\documentstyle[aps,prd,tighten,epsf]{revtex}
\begin{document}
\draft
\newcommand{\be}{\begin{equation}}
\newcommand{\ee}{\end{equation}}
\newcommand{\ben}{\begin{eqnarray}}
\newcommand{\een}{\end{eqnarray}}

\newcommand{\la}{{\lambda}}
\newcommand{\Om}{{\Omega}}
\newcommand{\ta}{{\tilde a}}
\newcommand{\bg}{{\bar g}}
\newcommand{\bh}{{\bar h}}
\newcommand{\si}{{\sigma}}
\newcommand{\th}{{\theta}}
\newcommand{\C}{{\cal C}}
\newcommand{\D}{{\cal D}}
\newcommand{\cA}{{\cal A}}
\newcommand{\cT}{{\cal T}}
\newcommand{\cO}{{\cal O}}
\newcommand{\eeo}{\cO ({1 \over E})}
\newcommand{\G}{{\cal G}}
\newcommand{\cL}{{\cal L}}
\newcommand{\T}{{\cal T}}
\newcommand{\M}{{\cal M}}

\newcommand{\p}{\partial}
\newcommand{\na}{\nabla}
\newcommand{\ssum}{\sum\limits_{i = 1}^3}
\newcommand{\dssum}{\sum\limits_{i = 1}^2}
\newcommand{\tal}{{\tilde \alpha}}

\newcommand{\tp}{{\tilde \phi}}
\newcommand{\tPhi}{\tilde \Phi}
\newcommand{\tpsi}{\tilde \psi}
\newcommand{\tim}{{\tilde \mu}}
\newcommand{\tr}{{\tilde \rho}}
\newcommand{\tir}{{\tilde r}}
\newcommand{\rp}{r_{+}}
\newcommand{\hr}{{\hat r}}
\newcommand{\rv}{{r_{v}}}
\newcommand{\dr}{{d \over d \hr}}
\newcommand{\dR}{{d \over d R}}

\newcommand{\hhf}{{\hat \phi}}
\newcommand{\hhM}{{\hat M}}
\newcommand{\hhQ}{{\hat Q}}
\newcommand{\hht}{{\hat t}}
\newcommand{\hhr}{{\hat r}}
\newcommand{\hhS}{{\hat \Sigma}}
\newcommand{\hhD}{{\hat \Delta}}
\newcommand{\hhm}{{\hat \mu}}
\newcommand{\hro}{{\hat \rho}}
\newcommand{\hhz}{{\hat z}}

\newcommand{\tD}{{\tilde D}}
\newcommand{\tB}{{\tilde B}}
\newcommand{\tV}{{\tilde V}}
\newcommand{\hT}{\hat T}
\newcommand{\tF}{\tilde F}
\newcommand{\tT}{\tilde T}
\newcommand{\hC}{\hat C}
\newcommand{\ep}{\epsilon}
\newcommand{\bep}{\bar \epsilon}
\newcommand{\ppp}{\varphi}
\newcommand{\Ga}{\Gamma}
\newcommand{\ga}{\gamma}
\newcommand{\hth}{\hat \theta}
\title{Reissner-Nordstr\"om Black Holes and Thick Domain Walls}

\author{Rafa{\l} Moderski}

\address{N.Copernicus Astronomical Center \protect \\
Polish Academy of Sciences, \protect \\
00-716 Warsaw, Bartycka 18, Poland \protect \\
moderski@camk.edu.pl}

\author{Marek Rogatko}

\address{Institute of Physics \protect \\
Maria Curie-Sklodowska University \protect \\
20-031 Lublin, pl.Marii Curie-Sklodowskiej 1, Poland \protect \\
rogat@tytan.umcs.lublin.pl \protect \\
 rogat@kft.umcs.lublin.pl}

\date{\today}
\maketitle
\smallskip
\bigskip
\begin{abstract}
We solve numerically equations of motion for real self-interacting
scalar fields in the background of Reissner-Nordstr\"om black hole and obtained 
a sequence of static axisymmetric solutions representing thick domain walls
charged black hole systems. In the case of extremal 
Reissner-Nordstr\"om black hole solution we find that there is a 
parameter depending on the black hole mass and the width of the domain wall
which constitutes the upper limit for the expulsion to occur.
\end{abstract}
\baselineskip=18pt
\section{Introduction}
Relics of cosmological phase transitions are considered to 
play an important role in cosmology \cite{vil94}. They are signs of the high-energy phenomena
which are beyond the range of contemporary accelerators. Such topological
defects as cosmic strings attracted great interests due to the motivation of the 
uniqueness theorems for black holes and Wheeler's conjecture {\it black holes have no hair}.
Black hole cosmic string configurations were studied from different points of view
\cite{dow92} revealing that cosmic string could thread it
or could be expelled from black hole. 
As far as the dilaton gravity was concerned it was 
also justified \cite{string,san00}
that a vortex could be treated by the remote observer 
as a hair on the black hole. 
On the other hand, for the extreme dilaton black hole one has 
always the expulsion of the Higgs field (the so-called {\it Meissner effect}).
The problem of vortices in de Sitter (dS) background was analyzed in Ref.\cite{ghe02a},
while the behaviour of Abelian Higgs vortex solutions in 
Schwarzschild anti de Sitter (AdS),
Kerr, Kerr-AdS and
Reissner-Nordstr\"om AdS was studied in Refs.\cite{deh02,ghe02b}. The vortex solution for
Abelian Higgs field Eqs. in the background of four-dimensional black
string was obtained in Ref.\cite{deh02b}.
\par
In order to investigate the behaviour of the domain walls or cosmic strings
configurations in curved spacetime a number of works use a thin wall
or string approximations, i.e., infinitely thin non-gravitating membrane
described by the Nambu-Goto Lagrangian.
Dynamics of scattering and capture process of an infinitely thin cosmic 
string in the spacetime
Schwarzshild black hole were elaborated in Ref.\cite{vil98}.
Christensen {\it et al.} \cite{chr98} revealed that there existed a family of 
infinitely thin walls
intersecting black hole event horizon. Then, the considerations were
applied to the scattering of inifitely long cosmic string by a
rotating black hole \cite{sna02}.
\par
Nowadays, the idea that the Universe is embedded in higher-dimensional
spacetime acquires much attention.
The resurgence of the idea is
motivated by the
possibility of resolving the hierarchy problem \cite{ran99}, i.e.,
the difference in magnitudes of the Planck scale and the electroweak
scales. Also the most promising candidate for a unified theory of Nature,
superstring theory predicts the existence of the so-called D-branes which in turn
renews the idea of {\it brane worlds}.
From this theory point of view the model of brane world stems from the idea 
presented by Horava and Witten \cite{hor96}. Namely, the strong coupling limit
of $E_{8} \times E_{8}$ heterotic string theory at low energy is
described by eleven-dimensional supergravity with eleventh dimension
compactified on an orbifold with $Z_{2}$ symmetry. The two boundaries
of spacetime are ten-dimensional planes, to which gauge theories are
confined. Next in Refs.\cite{wit96,luk99} it was argued 
that six of the eleven dimensions could be
consistently compactified and in the limit spacetime looked five-dimensional
with four-dimensional boundary brane.\\
Then, it is intriguing to
study the interplay between black holes and domain walls (branes).
Nowadays, this challenge
acquires more attention.
The problem of stability of a Nambu-Goto membrane in Reissner-Nordsr\"om
de Sitter (RN-dS) spacetime was studied
in Ref.\cite{hig01}, while the gravitationally interacting system 
of a thick domain wall
and Schwarschild black hole was considered in \cite{mor00,mor03}. 
Emparan { \it et al.} \cite{emp01}
elaborated the problem of a black hole on a topological domain wall.
In \cite{rog01} the dilaton black hole-domain wall system was studied 
analytically
and it was revealed that for the extremal dilaton black hole one had 
to do with the expulsion
of the scalar field from the black hole (the so-called {\it Meissner effect}). 
The numerical
studies of domain wall in the spacetime of dilaton black hole was 
considered in Ref.\cite{mod03},
where the thickness of the domain wall and a potential of the scalar 
field $\varphi^4$ and
sine-Gordon were taken into account.
\par
In our paper we try to provide some continuity with the previous works \cite{rog01,mod03}
concerning the behaviour of dilaton black hole domain wall systems,
namely we shall consider the thick domain wall in the background of RN black hole.
\par
The paper is organized as follows. Sec.II is devoted to the basic equations of the
considered problem. In Sec.III we studied the RN-domain wall system both for
$\varphi^4$ and sine-Gordon potential. We pay special attention to the extremal
RN black hole case and reveal that there is an expulsion of the domain wall
fields under certain conditions. In Sec.IV we concluded our investigations.

\section{Reissner-Nordstr\"om black hole domain wall system}
In our paper we shall consider the spherically symmetric solution 
of Einstein-Maxwell equations described by the
RN black hole metric of the form
\be
ds^2 = - \left ( 1 - {2 M \over r} + {Q^2 \over r^2}
\right ) dt^2 +
{d r^2 \over {\left ( 1 - {2 M \over r} + {Q^2 \over r^2}
\right )}} + r^2 (d \theta^2
+ \sin^2 \theta  d \ppp^2).
\ee
The metric is asymptotically flat in the sense that
the spacetime  
contains a data set
$(\Sigma_{end}, g_{ij}, K_{ij})$ with gauge fields such that 
$\Sigma_{end}$ is diffeomorphic to ${\bf R}^3$ minus a ball and the 
following asymptotic conditions are fulfilled:
\ben
\vert g_{ij}  - \delta_{ij} \vert + r \vert \p_{a}g_{ij} \vert
+ ... + r^k \vert \p_{a_{1}...a_{k}}g_{ij} \vert +
r \vert K_{ij} \vert + ... + r^k \vert \p_{a_{1}...a_{k}}K_{ij} \vert
\le {\cal O}\bigg( {1\over r} \bigg), \\
\vert F_{\alpha \beta} \vert + r \vert \p_{a} F_{\alpha \beta} \vert
+ ... + r^k \vert \p_{a_{1}...a_{k}}F_{\alpha \beta} \vert
\le {\cal O}\bigg( {1 \over r^2} \bigg).
\een
RN black hole spacetime will be the background metric on which
we shall consider the domain wall equations of motion
built of a self-interacting
scalar field. 
In our investigations we use
a general matter Lagrangian with real Higgs field and the
symmetry breaking potential of the form as follows:
\be
{\cal L}_{dw} = - {1 \over 2} \na_{\mu} \varphi \na^{\mu} \varphi
- U(\varphi),
\ee
where $\varphi$ real Higgs field and the
symmetry breaking potential $U(\varphi)$
has a discrete set of
degenerate minima.  The energy-momentum tensor for scalar field
may be written as follows:
\be
T_{ij}(\varphi) = - {1 \over 2} g_{ij} \na_{m} \varphi \na^{m} \varphi
- U(\varphi) g_{ij} + \na_{i} \varphi \na_{j} \varphi.
\label{ten}
\ee
It is turned out that 
for the convenience we can scale out parameters via transformation $ X =
{\varphi / \eta}$ and $\ep = 8 \pi G \eta^2$ \cite{emp01}.  The parameter $\ep$ is
responsible for 
the gravitational strength and it is also connected with the
gravitational interaction of the considered Higgs field. One can also define 
the potential $V(X) =
{U(\varphi) \over V_{F}}$, where $V_{F} = \lambda \eta^4$.
All these quantities
enable us to rewrite the Lagrangian for the domain wall fields in a more
suitable form. Namely, it now reads
\be
8 \pi G {\cal L}_{dw} = - {\ep \over w^2} \bigg[
w^2 {\na_{\mu} X \na^{\mu} X \over 2} + V(X) \bigg],
\label{dw}
\ee
where $w = \sqrt{{\ep \over 8 \pi G V_{F}}}$ represents the inverse
mass of the scalar after symmetry breaking. It characterizes
the width of the wall defect.
Just using Eq.(\ref{dw}) we arrive at the following expression for $X$ field:
\be
\na_{\mu} \na^{\mu} X - {1 \over w^2}{ \p V \over \p X} = 0.
\label{mo}
\ee
As in Ref.\cite{mod03} we shall take into account
for two cases of potentials with a discrete set
of degenerate minima, namely
the $\varphi^4$ potential described by the relation
\be
U_1(\varphi) = {\lambda \over 4} (\varphi^2 - \eta^2)^2,
\label{phi4}
\ee
and the sine-Gordon potential in the form of
\be
U_2(\varphi) = \lambda \eta^4 \left [ 1 + \cos(\varphi/\eta) \right ].
\label{sinG}
\ee

\section{Boundary conditions}
\subsection{Boundary conditions for $\varphi^4$ potential}
As we mentioned RN black hole spacetime is asymptotically
flat, thus the asymptotic boundary solution of equation of motion
for potential (\ref{phi4}) will be the solution of equation of motion in flat
spacetime. It reads as
\be
\varphi_1(z) = \eta \tanh ( \sqrt{\lambda/2} \eta z),
\label{solphi4}
\ee
or rewritten in our units implies the following:
\be
X(r,\theta) = \tanh \bigg( {r \cos \theta \over \sqrt{2} w }\bigg )
\ee
Thus,
the equation of motion (\ref{mo}) for the scalar field $X$ 
in the considered background has the form of
\be
{1 \over r^2 }
\p_{r} \bigg[ \big ( r^2 - 2 M r + Q^2 \big)
\p_{r} X \bigg] + 
{1 \over r^2 \sin \theta }
\p_{\theta} \bigg[ \sin \theta \p_{\theta} X \bigg] - {1 \over w^2}
X(X^2-1) = 0.
\label{motphi4}
\ee
and the energy density yields
\be
E = \bigg[
- {1 \over 2} \big( \p_r X \big)^2 \bigg( 1 - {2M \over r} + {Q^2 \over r^2} \bigg)
-{ 1 \over 2} \big( \p_\theta X \big)^2
{1 \over r^2 }
 \bigg] {w^2} - {1 \over 4} (X^2-1)^2.
\label{enphi4}
\ee
On the horizon of the RN black hole relation (\ref{motphi4}) gives 
the following boundary condition:
\be
{\rp - r_{-} \over \rp^2} \p_r X  \big|_{r = \rp} = - {1 \over \rp^2
\sin \theta} \p_\theta \left [ \sin \theta \p_\theta
X \right ] + {1 \over w^2}
X(X^2-1),
\label{bhor}
\ee
where $\rp = M + \sqrt{M^2 - Q^2}$ and $r_{-} = M - \sqrt{M^2 - Q^2}$ are
respectively the outer and the inner horizon of the RN black hole.\\
As in Ref.\cite{mod03} 
we shall investigate
only the case when the core of the wall is
located in the equatorial plane $\theta = \pi/2$ of the black hole, we
impose the Dirichlet boundary condition at the equatorial plane
\be
X \big|_{\theta = \pi/2} = 0,
\label{beq}
\ee
as well as
the regularity of the scalar field on the symmetric axis requires the
Neumann boundary condition on the z-axis:
\be
{\p X \over \p \theta} \bigg|_{\theta=0} = 0.
\label{bz}
\ee
For the asymptotical flatness of the RN black hole solution we should
have far from the black hole the flat spacetime solution 
(\ref{solphi4}). But we are limited by our computational grid which is finite,
then we require the following boundary condition ought to be satisfy
at the boundary of the grid
\be
X \big|_{r=r_{max}} = \tanh \bigg({ r_{max} \cos \theta \over \sqrt{2} w} \bigg).
\label{brmax}
\ee

\subsection{Boundary conditions for the sine-Gordon potential.}
As far as
the sine-Gordon potential (\ref{sinG}) is concerned the flat spacetime solution
is given by
\be
\varphi_2(z) = \eta \left \{ 4 \arctan \left [ \exp (\sqrt{\lambda}
\eta z ) \right ] - \pi \right \},
\label{sinGsol}
\ee
or in our units this is equivalent to following:
\be
X(r,\theta) = 4 \arctan \bigg [ \exp \bigg({r \cos \theta \over w}\bigg) \bigg ] - \pi.
\label{XsinG}
\ee
In this case $V(X) = 1 + \cos(X)$ and $\p V / \p X = - \sin(X)$, and
thus the equation of motion (\ref{mo}) takes the form
\be
{1 \over r^2} \bigg[
\p_{r} \bigg( r^2 - 2 M r + Q^2 \bigg)
\p_{r} X \bigg] + 
{1 \over r^2 \sin \theta }
\p_{\theta} \bigg[ \sin \theta \p_{\theta} X \bigg] + 
{1 \over w^2} \sin(X) = 0,
\label{motsinG}
\ee
while the energy density is given by the relation
\be
E = \bigg[
- {1 \over 2} \big( \p_r X \big)^2 \bigg( 1 - {2M \over r} + {Q^2 \over r^2} \bigg)
-{ 1 \over 2} \big( \p_\theta X \big)^2
{1 \over r^2}
 \bigg] {w^2} - 1 - \cos(X).
\label{ensinG}
\ee
On the horizon from Eq.(\ref{motsinG}) one has the boundary condition as follows:
\be
{\rp - r_{-} \over \rp^2}
\p_r X  \big|_{r= \rp} = - {1 \over \rp^2
\sin \theta} \p_\theta \left [ \sin \theta \p_\theta
X \right ] - {1 \over w^2} \sin(X).
\label{bhorsinG}
\ee
Of course, this must be accompanied by the Dirichlet boundary conditions at the equatorial
plane of the black hole (\ref{beq}), and the Neumann
boundary condition on the symmetry axis (\ref{bz}),
and at the outer edge of the grid
\be
X \big|_{r=r_{max}} = 4 \arctan \bigg[ \exp \bigg({r_{max}\cos \theta \over w}\bigg)
\bigg ] - \pi.
\ee
%

\par
Another interesting problem
that should attract attention to is the case
of extremal RN black hole domain wall configuration.
As was revealed analitycally in Ref.\cite{emp01} that for some range of parameters we should have 
the expulsion of the domain wall, contrary to the situation which takes
places in extremal dilaton black holes \cite{rog01,mod03}.
Simple arguments given in Ref.\cite{emp01}
it was revealed that for
at least very small extremal black hole sitting inside the domain wall,
the black hole would expel it. The mass bound for the extremal RN black hole 
for the expulsion was also found. For $w = 1$ the black hole mass have to satisfy
the condition
$M^2<{1 \over 2}$.
\par
When $\rp = r_{-}$ we have to do with the extremal RN black hole and 
it implies the condition
$Q = M$. For this case the metric has the following form:
\be
ds^2 = - \left ( 1 - {M \over r}  
\right )^2 dt^2 +
{d r^2 \over {\left ( 1 - {M \over r}
\right )^2}} + r^2 (d \theta^2
+ \sin^2 \theta  d \ppp^2).
\ee
Equation of motion for the domain wall's scalar fields 
in the background of RN extremal black hole yields
\be
{1 \over r^2 }
\p_{r} \bigg[ \big ( r - M \big)^2
\p_{r} X \bigg] + 
{1 \over r^2 \sin \theta }
\p_{\theta} \bigg[ \sin \theta \p_{\theta} X \bigg] = {1 \over w^2}{\p V \over \p X},
\label{eks}
\ee
while the relation for energy implies the following:
\be
E = \bigg[
- {1 \over 2} \big( \p_r X \big)^2 \bigg( 1 - {M \over r} \bigg)^2
-{ 1 \over 2} \big( \p_\theta X \big)^2
{1 \over r^2 }
 \bigg] {w^2} - V(X).
\label{eks1}
\ee

\section{Numerical calculations}
\subsection{Reissner-Nordstr\"om  black hole}
\begin{figure}
\begin{center}
\leavevmode
\epsfxsize=440pt
\epsfysize=540pt
\epsfbox{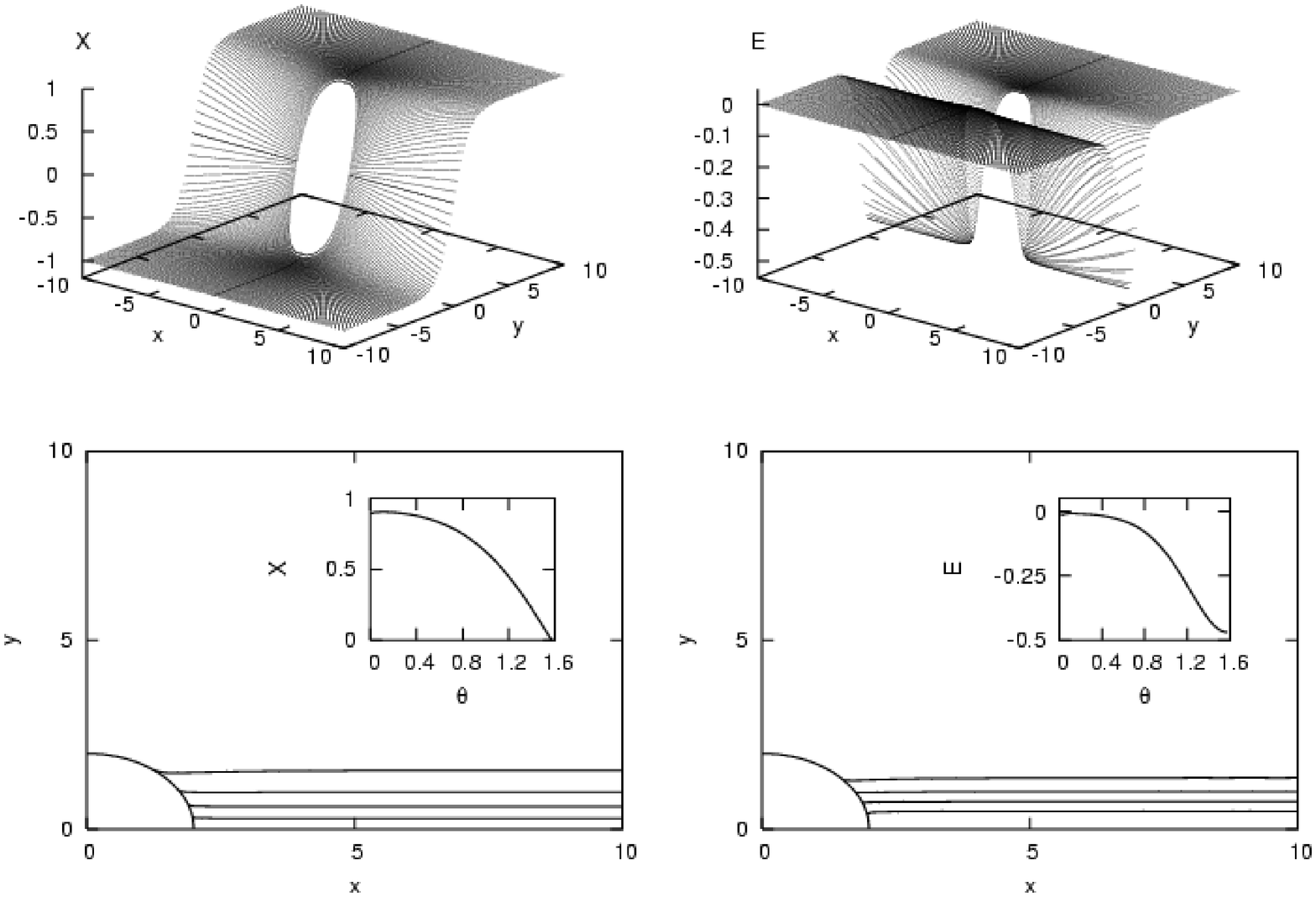}
\end{center}
\caption{
  The field X (left panels) and the energy E (right panels)
  for the $\phi^4$ potential and the Reissner-Nordstr\"om
  black hole. Isolines on bottom panels are drawn for $0.2$, $0.4$,
  $0.6$ and $0.8$ for the field X and for $-0.1$, $-0.2$, $-0.3$ and
  $-0.4$ for the energy. Inlets in bottom plots show the value of the
  fields on the black hole horizon. Black hole has $M=1.0$, $Q=0.1$
  and the domain width is $w=1.0$.}
\label{fig1a}
\end{figure}

\begin{figure}
\begin{center}
\leavevmode
\epsfxsize=440pt
\epsfysize=540pt
\epsfbox{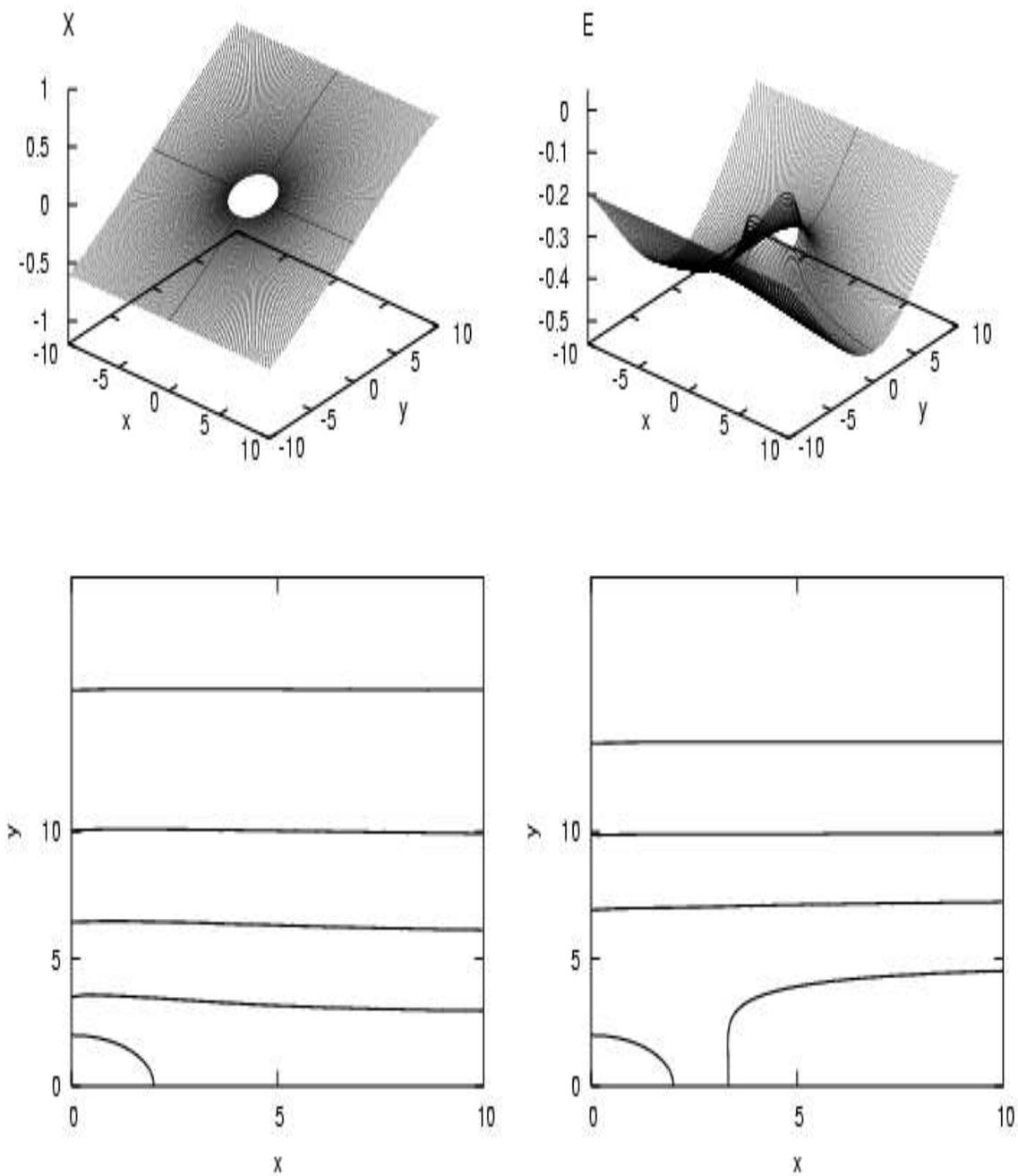}
\end{center}
\caption{
  The field X (left panels) and the energy E (right panels)
  for the $\phi^4$ potential and the Reissner-Nordstr\"om
  black hole. Isolines on bottom panels are drawn for $0.2$, $0.4$,
  $0.6$ and $0.8$ for the field X and for $-0.1$, $-0.2$, $-0.3$ and
  $-0.4$ for the energy. Black hole has $M=1.0$, $Q=0.1$
  and the domain width is $w = 10$.}
\label{fig1b}
\end{figure}

\begin{figure}
\begin{center}
\leavevmode
\epsfxsize=440pt
\epsfysize=540pt
\epsfbox{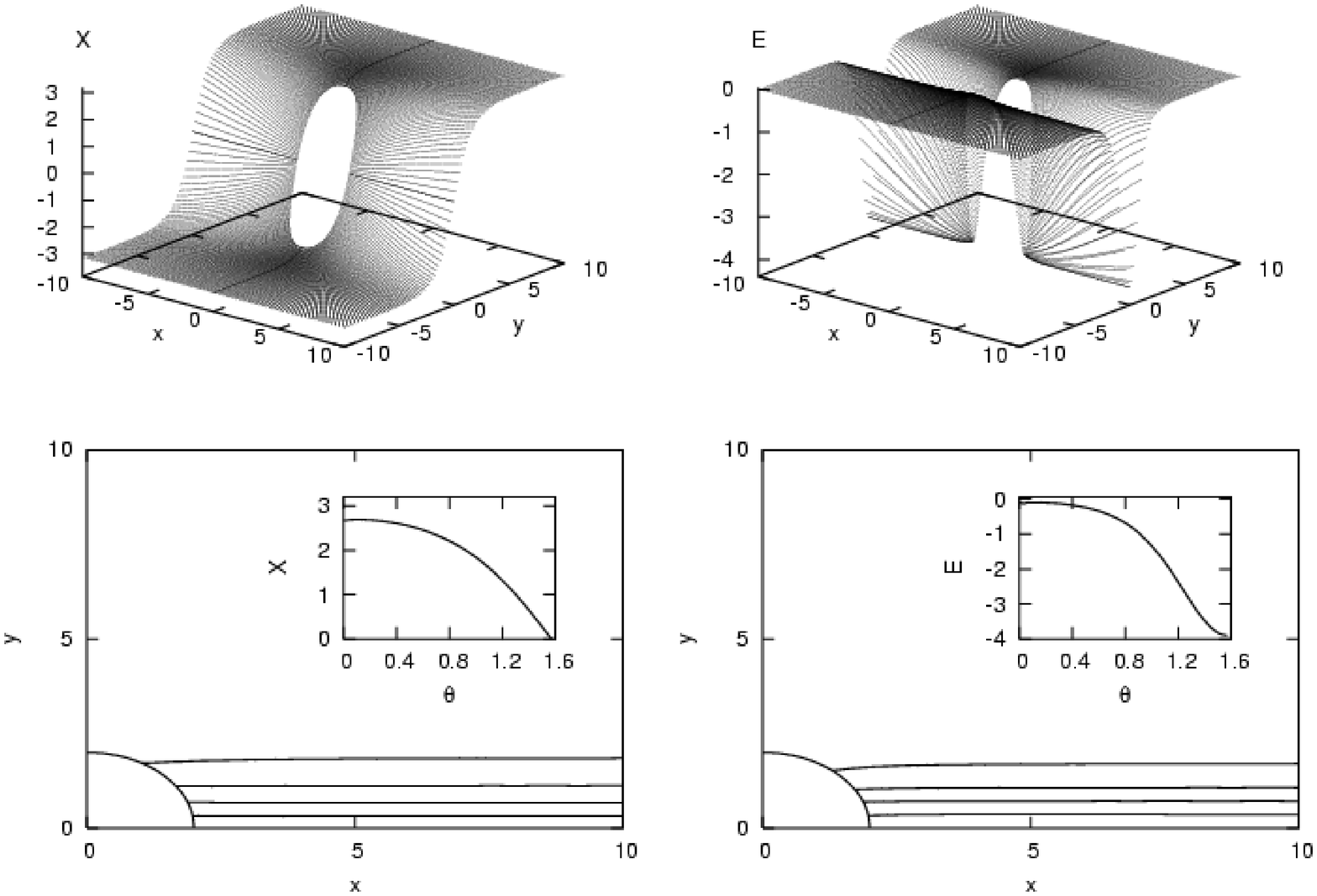}
\end{center}
\caption{
  The field X (left panels) and the energy E (right panels)
  for the sine-Gordon potential and the Reissner-Nordstr\"om
  black hole. Isolines on bottom panels are drawn for $0.2\pi$, $0.4\pi$,
  $0.6\pi$ and $0.8\pi$ for the field X and for $-0.5$, $-1.5$, $-2.5$ and
  $-3.5$ for the energy. Inlets in bottom plots show the value of the
  fields on the black hole horizon. Black hole has $M=1.0$, $Q=0.1$
  and the domain width is $w=1.0$.}
\label{fig1c}
\end{figure}

\begin{figure}
\begin{center}
\leavevmode
\epsfxsize=440pt
\epsfysize=540pt
\epsfbox{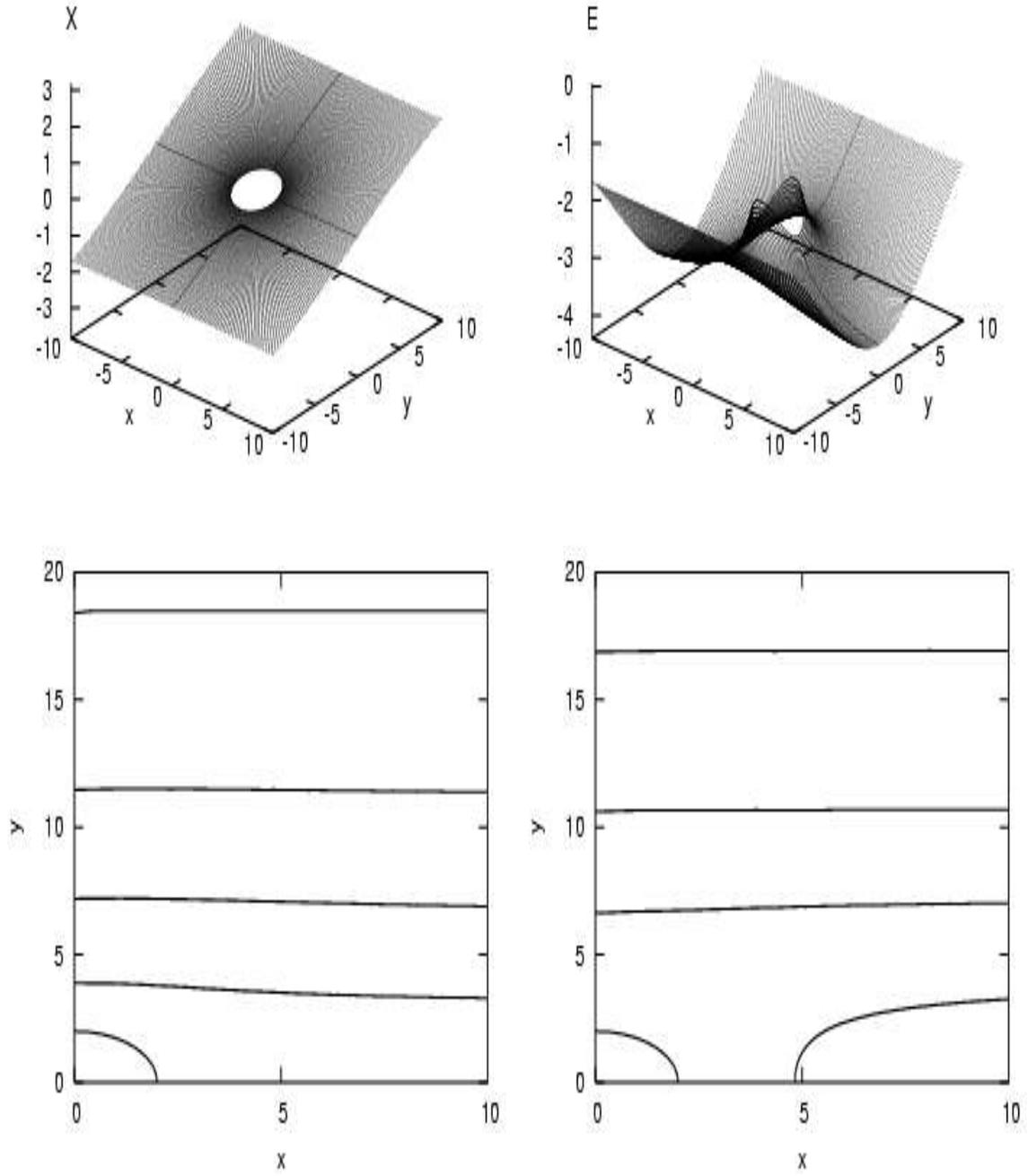}
\end{center}
\caption{ 
  The field X (left panels) and the energy E (right panels)
  for the sine-Gordon potential and the Reissner-Nordstr\"om black
  hole. Isolines on bottom panels are drawn for $0.2\pi$, $0.4\pi$,
  $0.6\pi$ and $0.8\pi$ for the field X and for $-0.5$, $-1.5$, $-2.5$
  and $-3.5$ for the energy. Black hole has $M=1.0$, $Q=0.1$ and the
  domain width is $w=10$.}
\label{fig1d}
\end{figure}

The method of solving equations of motion for the scalar fields $X$ is the
same as in Ref.\cite{mod03}. We applied the modified overrelaxation
method in order to take into account the boundary conditions on the
considered black hole horizon. Eqs. of motion will be solved on a
uniformly spaced polar grid $(r_{i}, \theta_{i})$ with the boundaries
at $r_{min} = r_{+}$, the outer radius will be $r_{max} \gg r_{+}$ (we
take usually $r_{max} = 20 r_{+}$ ).  The angle $\theta$ changes from
$0$ to $\pi/2$.  Then, one approximates the derivatives with
finite-difference expressions on the grid.  The rest of the solution
for scalar field $X$ will be obtained from the symmetry condition.
The example solutions of the equation of motion for different
parameters and field configurations are shown on Figs~\ref{fig1a},
\ref{fig1b}, \ref{fig1c}, and \ref{fig1d}.

\subsection{An extreme Reissner-Nordstr\"om black hole}
\begin{figure}
\begin{center}
\leavevmode
\epsfxsize=440pt
\epsfysize=540pt
\epsfbox{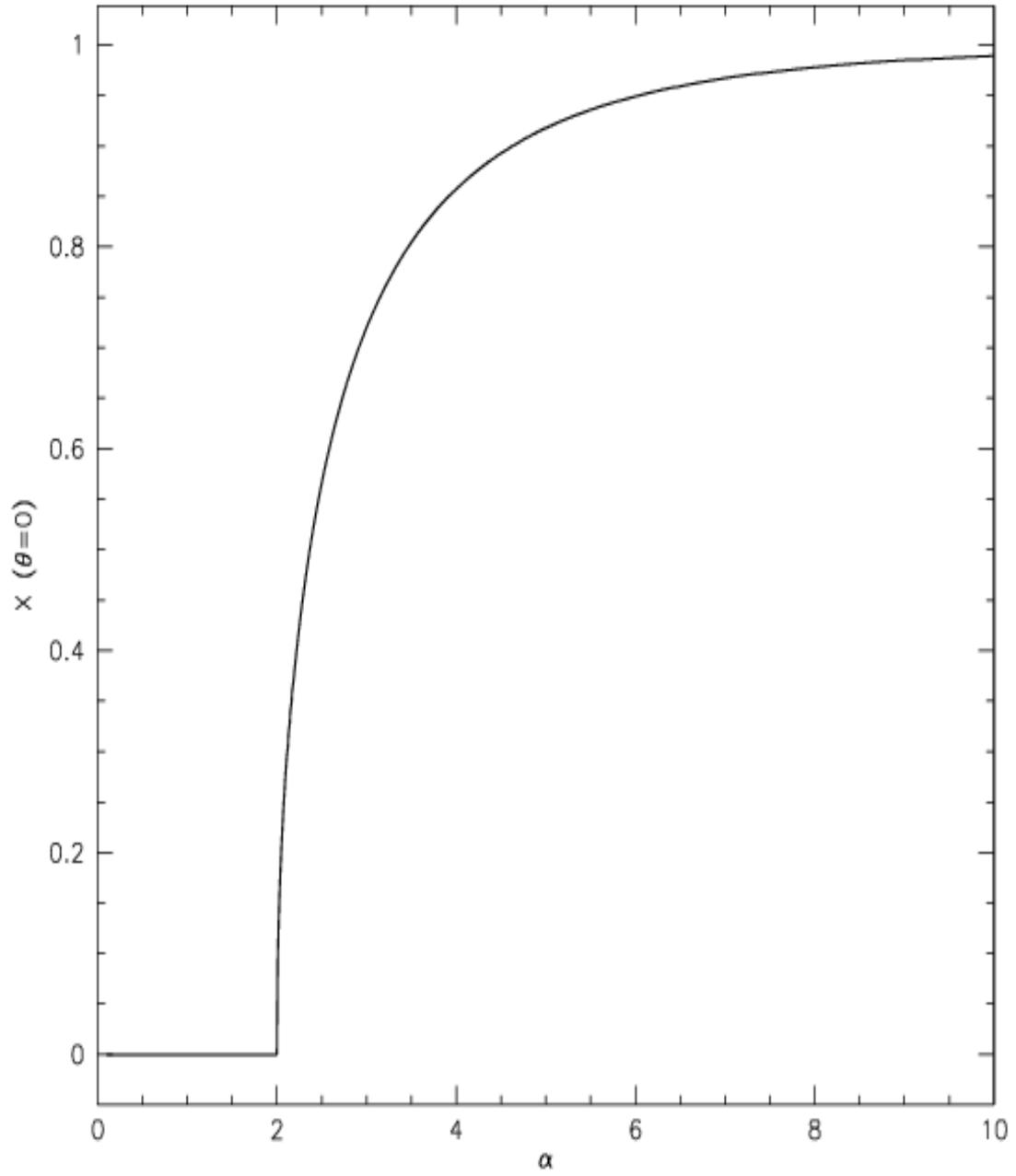}
\end{center}
\caption{ 
  Value of the field X on the black hole axis ($\theta=0$) as
  a function of the parameter $\alpha$ for extreme
  Reissner-Nordstr\"om black hole and $\phi^4$ potential.}
\label{fig2a}
\end{figure}

\begin{figure}
\begin{center}
\leavevmode
\epsfxsize=440pt
\epsfysize=540pt
\epsfbox{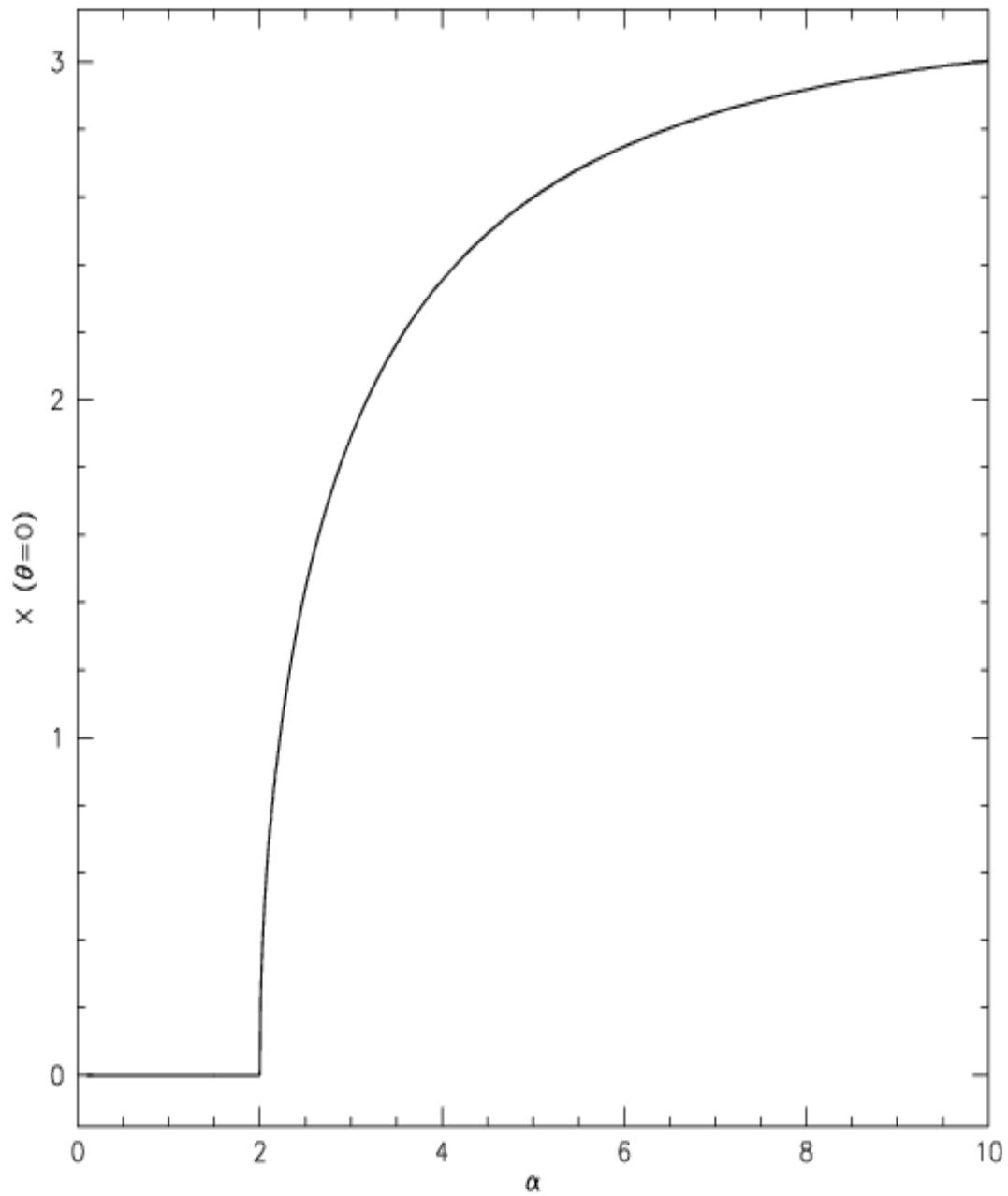}
\end{center}
\caption{ 
  Same as Fig.~\ref{fig2a}, but for the sine-Gordon potential.}
\label{fig2b}
\end{figure}

For the extremal RN  black hole scalar fields on the horizon decouple
from the rest of
the grid and equation of motion (\ref{eks}) becomes an ordinary
differential equation:
\be
\partial_{\theta\theta} X + {1 \over \tan \theta} \partial_{\theta} X - 
{r_+^2 \over w^2} {\partial V \over \partial X} = 0.
\label{hor}
\ee
For the reasons given in \cite{dow92,mod03} one must carefully examine
the solution on the horizon. The solution $X=0$, which
corresponds to the expulsion of the field from the black hole (the
{\it Meissner effect}) always solves relation (\ref{hor}). But sometimes
other solution is possible and energetically more favourable. The previous
studies envisage (\cite{dow92,mod03}) that the relative
size of the domain wall 
compared
to the size of the black hole becomes a factor
which determines the overall behaviour of the field on the horizon.
For this reason we parameterize equation (\ref{hor}) with parameter
$\alpha \equiv {r_+^2 \over w^2}$ and solve (\ref{hor}) numerically
using two point boundary relaxation method (\cite{num}).
Fig.~\ref{fig2a} shows the dependence of the scalar field 
$X$
on the black hole
z-axis as a function of the parameter $\alpha$ for the $\varphi^4$
potential.  Fig.~\ref{fig2b} depicts the same dependence for the
sine-Gordon potential.  The difference 
of the scalar field
behaviours is clearly visible.
For the value $\alpha<2.0$ we have the field expulsion, while for $\alpha>2.0$
the field pierces the horizon of the extremal RN black hole.  The value
of the critical parameter is the same in both cases, which is easily
understood when one considers the series expansion of the $\partial
V/\partial X$ around zero and neglects higher terms.

\begin{figure}
\begin{center}
\leavevmode
\epsfxsize=440pt
\epsfysize=540pt
\epsfbox{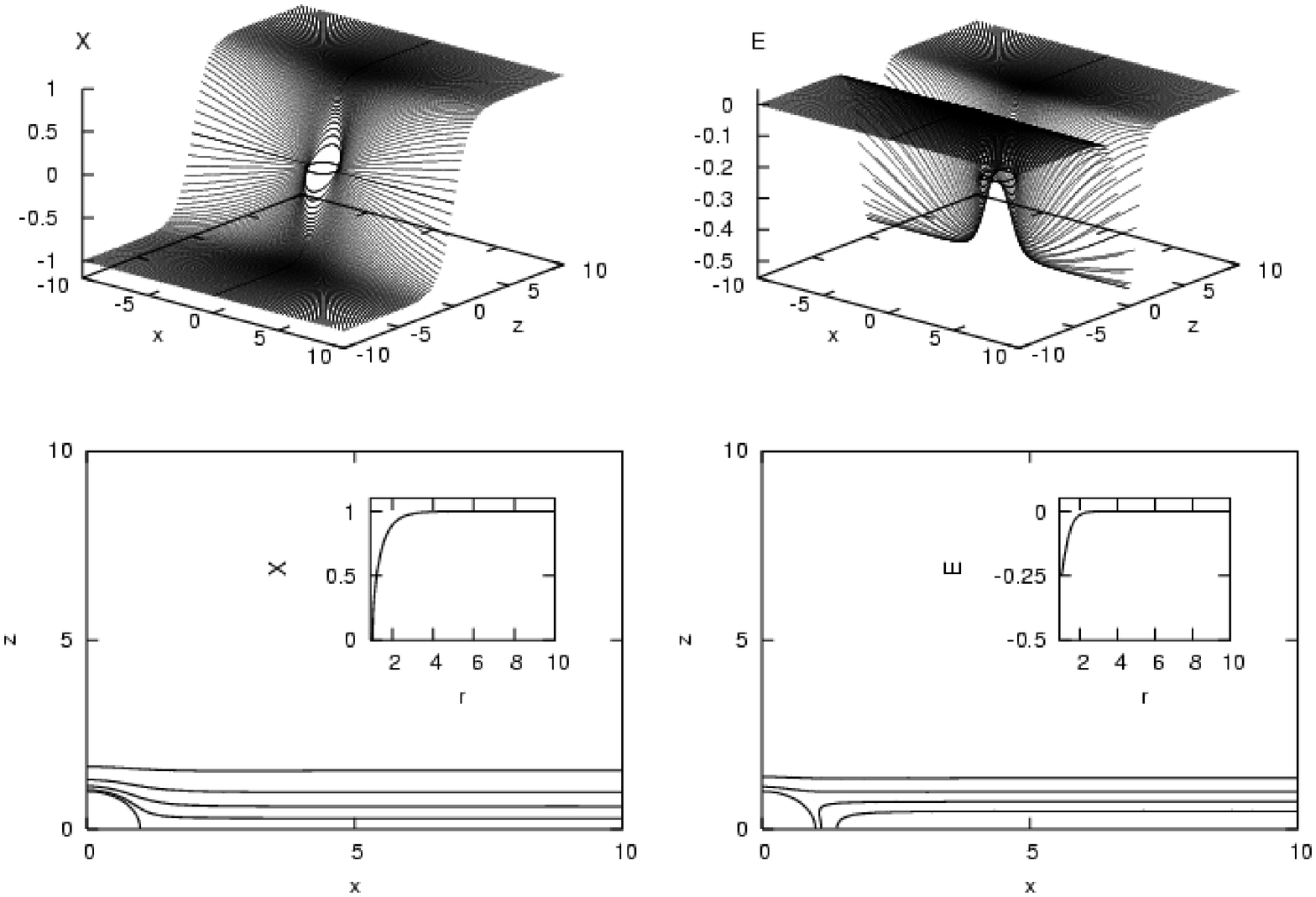}
\end{center}
\caption{
  The field X (left panels) and the energy E (right panels)
  for the $\phi^4$ potential and the extreme Reissner-Nordstr\"om
  black hole. Isolines on bottom panels are drawn for $0.2$, $0.4$,
  $0.6$ and $0.8$ for the field X and for $-0.1$, $-0.2$, $-0.3$ and
  $-0.4$ for the energy. Inlets in bottom plots show the value of the
  fields on the black hole horizon. Black hole has $M=1.0$, $Q=1.0$
  and the domain width is $w=1.0$, so the parameter $\alpha = 1.0$.}
\label{fig3a}
\end{figure}

\begin{figure}
\begin{center}
\leavevmode
\epsfxsize=440pt
\epsfysize=540pt
\epsfbox{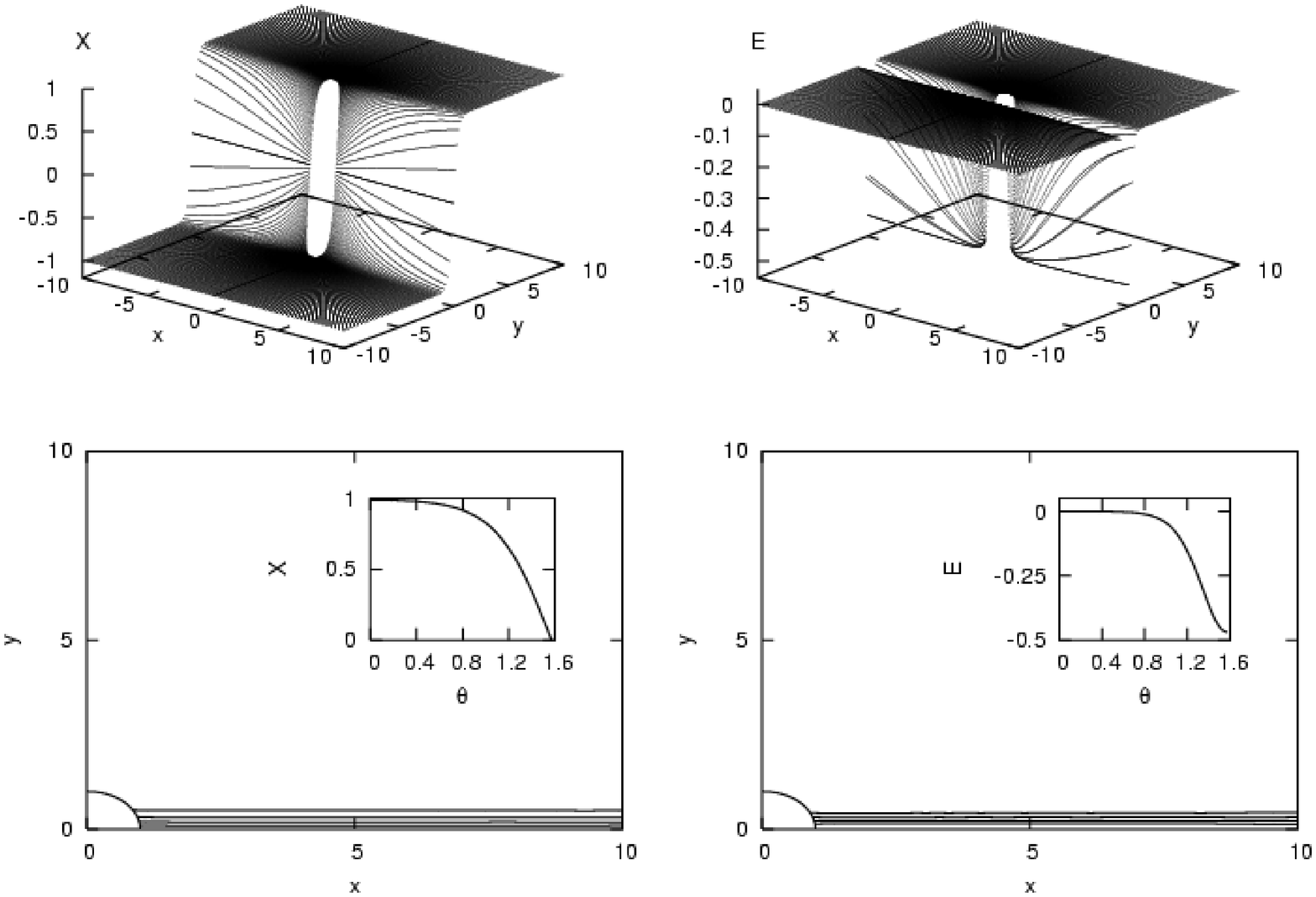}
\end{center}
\caption{
  The field X (left panels) and the energy E (right panels)
  for the $\phi^4$ potential and the extreme Reissner-Nordstr\"om
  black hole. Isolines on bottom panels are drawn for $0.2$, $0.4$,
  $0.6$ and $0.8$ for the field X and for $-0.1$, $-0.2$, $-0.3$ and
  $-0.4$ for the energy. Inlets in bottom plots show the value of the
  fields on the black hole horizon. Black hole has $M=1.0$, $Q=1.0$
  and the domain width is $w=\sqrt{0.1}$, so the parameter $\alpha = 10.0$.}
\label{fig3b}
\end{figure}

\begin{figure}
\begin{center}
\leavevmode
\epsfxsize=440pt
\epsfysize=540pt
\epsfbox{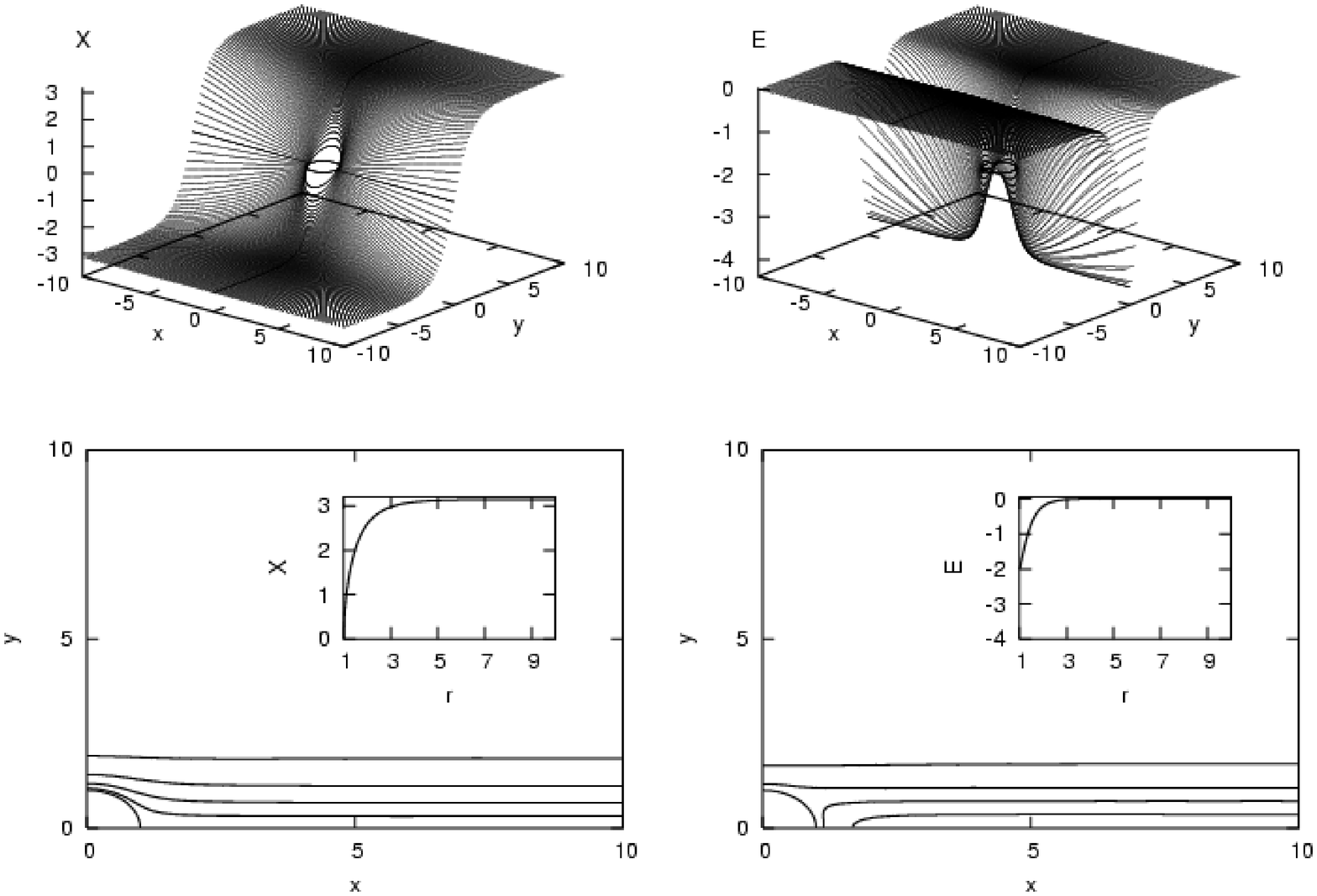}
\end{center}
\caption{
  The field X (left panels) and the energy E (right panels)
  for the sine-Gordon potential and the extreme Reissner-Nordstr\"om
  black hole. Isolines on bottom panels are drawn for $0.2\pi$, $0.4\pi$,
  $0.6\pi$ and $0.8\pi$ for the field X and for $-0.5$, $-1.5$, $-2.5$ and
  $-3.5$ for the energy. Inlets in bottom plots show the value of the
  fields on the black hole horizon. Black hole has $M=1.0$, $Q=1.0$
  and the domain width is $w=1.0$, so the parameter $\alpha = 1.0$.}
\label{fig3c}
\end{figure}

\begin{figure}
\begin{center}
\leavevmode
\epsfxsize=440pt
\epsfysize=540pt
\epsfbox{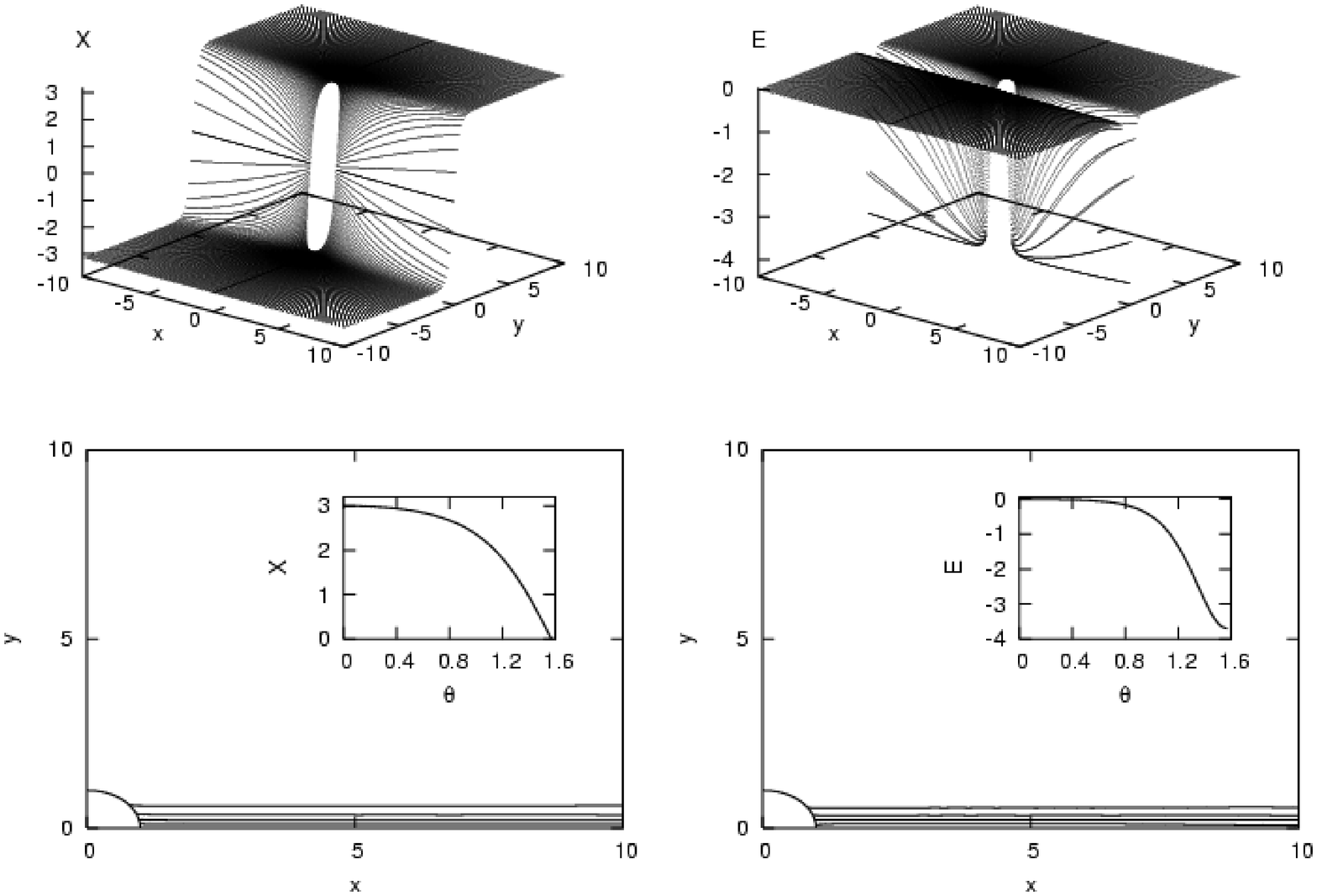}
\end{center}
\caption{
  The field X (left panels) and the energy E (right panels)
  for the sine-Gordon potential and the extreme Reissner-Nordstr\"om
  black hole. Isolines on bottom panels are drawn for $0.2\pi$, $0.4\pi$,
  $0.6\pi$ and $0.8\pi$ for the field X and for $-0.5$, $-1.5$, $-2.5$ and
  $-3.5$ for the energy. Inlets in bottom plots show the value of the
  fields on the black hole horizon. Black hole has $M=1.0$, $Q=1.0$
  and the domain width is $w=\sqrt{0.1}$, so the parameter $\alpha = 10.0$.}
\label{fig3d}
\end{figure}

Figs~\ref{fig3a}, \ref{fig3b}, \ref{fig3c}, \ref{fig3d} show examples
of the field configuration for various combinations of the $\alpha$
parameter and the adequate potentials.
\par
Summing it all up we conclude that the numerical studies of the scalar domain wall
field in the vicinity of RN black hole reveal that for the extremal case one has not 
always expulsion of the fields from the black hole. This phenomenon
depends on the value of the black hole radius $r_{+}$ and the width
of the domain wall. If the parameter $\alpha<2.0$ we have expulsion of the scalar
field $X$ while for $\alpha>2.0$ the scalar fields penetrate the extremal RN
black hole horizon. The domain wall extremal RN black hole system was studied
analytically in Ref.\cite{emp01}. By simple argument it was revealed that at least small
extremal RN black holes sitting inside the domain wall would expel it.
Then a more precise bound on the mass of the considered black hole was established.
For $M^2<1/2$ the defect's field cease to penetrate the black hole horizon, but for
$M^2>1/2$ one has no such expelling. In Ref.\cite{emp01} the considerations
were conducted for the case of the domain wall's width equaled to $1$.
\par
On the contrary, in the case of extremal dilaton black hole domain wall system
it was shown that one had always expulsion of the scalar field \cite{rog01,mod03}.
It turned out that the simplest generalization of Einstein-Maxwell theory by
adding massless dilaton field dramatically changed the structure of the extremal
black hole. The similar situation takes place when one consideres
anothe topological defect, i.e., cosmic string. For extremal dilaton black holes 
we observe always expulsion of the fields \cite{string}.

\section{Conclusions}
In our paper we elaborated the problem of thick domain walls in the background
of a spherically symmetric solution of Einstein-Maxwell field equations.
The domain wall equations of motion were built of a self-interacting
real scalar fields with the symmetry breaking potential having a discrete
set of degenerate minima. Using the modified overrelaxation method modified
in order to comprise the boundary conditions of the scalar field $X$
on the black hole horizon we solved numerically Eqs. of motion for
$\varphi^4$ and sine-Gordon potentials. All our solutions depended on
the parameter $w = 1/\sqrt{\lambda} \eta$ which was responsible for
the domain wall thickness. We revealed the systems of axisymmetric
scalar field configurations representing thick domain wall in the nearby
of RN black hole for both kinds of potentials. We paid special attention
to the extremal RN black hole solution. We parametrized the equations of motion
with the parameter $\alpha = {r_{+}^2 \over w^2}$ and solved it by means
of the two-point boundary relaxation method.
For $\alpha<2$ one observed the complete expulsion of the scalar field $X$
from the extremal RN black hole (the so-called {\it Meissner effect}), while
for $\alpha>2$ the domain wall's field penetrated the extremal
RN black hole horizon. This behaviour is in contrast to
the behaviour of scalar field in the background of the extremal dilaton
black hole, where we have a complete expulsion of the scalar fields.


\vspace{2cm}
\noindent
{\bf Acknowledgements:}\\
M.R. was supported in part by KBN grant No. 2 P03B 124 24.

\pagebreak





\end{document}